\documentclass[aps,preprint,tightenlines,showpacs]{revtex4}%
\usepackage{amsfonts}
\usepackage{amsmath}
\usepackage{amssymb}
\usepackage{graphicx}%
\usepackage{pifont}
\usepackage{wasysym}
\usepackage{multirow}
\providecommand{\U}[1]{\protect\rule{.1in}{.1in}}

\begin{document}
\title{Benchmark calculations for elastic fermion-dimer scattering}
\author{Shahin Bour$^{a}$, H.-W. Hammer$^{a}$, Dean~Lee$^{b}$, Ulf-G.~Mei{\ss }%
ner$^{a,c}$}
\affiliation{$^{a}$Helmholtz-Institut f\"{u}r Strahlen- und Kernphysik and Bethe Center for
Theoretical Physics, Universit\"{a}t Bonn, D-53115 Bonn, Germany
\linebreak$^{b}$Department of Physics, North Carolina State University,
Raleigh, NC 27695, USA \linebreak$^{c}$Institut f\"{u}r Kernphysik, Institute
for Advanced Simulation, JARA-HPC and J\"{u}lich Center for Hadron Physics,
Forschungszentrum J\"{u}lich, D-52425 J\"{u}lich, Germany}

\begin{abstract}
We present continuum and lattice calculations for elastic scattering between a
fermion and a bound dimer in the shallow binding limit. \ For the continuum
calculation we use the Skorniakov-Ter-Martirosian (STM) integral equation to
determine the scattering length and effective range parameter to high
precision. \ For the lattice calculation we use the finite-volume method of
L\"{u}scher. \ We take into account topological finite-volume corrections to
the dimer binding energy which depend on the momentum of the dimer. \ After
subtracting these effects, we find from the
lattice calculation $\kappa a_{fd} = 1.174(9)$ and $\kappa 
r_{fd} = -0.029(13)$. \ These results agree well with the continuum values
$\kappa a_{fd}=1.17907(1)$ and $\kappa r_{fd}=-0.0383(3)$
obtained from the STM equation. \ We discuss applications to cold
atomic Fermi gases, deuteron-neutron scattering in the spin-quartet channel,
and lattice calculations of scattering for nuclei and hadronic molecules at
finite volume.

\end{abstract}

\maketitle

\section{Introduction}

In this paper, we perform the first benchmark of finite-volume lattice methods
for the low-energy scattering of composite objects. \ Our results will have
immediate applications to lattice studies of elastic neutron-nucleus scattering.
 \ In the analysis presented here we consider
scattering between a fermion and a bound dimer composed of two fermions. \ In
order to test the precision of our lattice calculations, we also repeat the
same calculations using the Skorniakov-Ter-Martirosian (STM) integral 
equation \cite{Skorniakov:1957}. 
\ Along the way we also provide the most
accurate calculation to date for the fermion-dimer effective range parameter.
\ Some of the results presented were summarized in a letter publication
\cite{Bour:2011ef}, and we present the full details of the calculations here.

We consider two component fermions. \ We will refer to the two fermion
components as spin up and spin down and consider the case when the masses are
equal, $m_{\uparrow}=m_{\downarrow}$. \ We assume finite-range attractive
interactions and consider the universal shallow binding limit. \ If $R$ is the
range of the interactions and $\kappa$ is the binding momentum of the dimer,
then the shallow binding limit corresponds to $\kappa R\rightarrow0$.

We note that much of the literature on universal three-body systems has
focused on the Efimov effect \cite{Efimov:1970} for
three bosons, three-component fermions, unequal mass fermions, or mixed
Bose-Fermi systems \cite{Kraemer:2006,Chin:2010,Braaten:2006,Platter:2009,
Helfrich:2010}. \ For equal mass two-component fermions,
however, there are no short-distance three-body instabilities such as the 
Thomas collapse \cite{Thomas:1935}. \ Hence there are no relevant
momentum scales other than the dimer binding momentum $\kappa$, and all
low-energy scattering parameters can be expressed as dimensionless constants
times the corresponding power of $\kappa$. \ In the shallow binding limit
$\kappa$ is the same as the reciprocal of the fermion-fermion scattering length.

There have been numerous calculations of the fermion-dimer scattering length.
\ The first goes back to the early work of Skorniakov and Ter-Martirosian who
found $\kappa a_{fd}\approx1.2$ \cite{Skorniakov:1957}. \ An overview of the
Skorniakov-Ter-Martirosian (STM) integral equation method will be presented in
our discussion below. \ The same value $\kappa a_{fd}\approx1.2$ has been
confirmed several times using integral equations
\cite{Petrov:2003,Petrov:2005}. \ A value of $\kappa a_{fd}\approx1.11$ was
obtained using an epsilon expansion in dimensions \cite{Rupak:2006jj}. \ A
recent correlated Gaussian expansion calculation obtained $\kappa
a_{fd}\approx1.18(1)$ \cite{vonStecher:2008}. \ This agrees with
integral equation studies which found $\kappa a_{fd}\approx1.18$
\cite{Levinsen:2008,Levinsen:2011} and $\kappa a_{fd}\approx1.1790662349$
\cite{Tan:2008}.

The fermion-dimer results at shallow binding should approximately describe
neutron-deuteron scattering in the spin-quartet channel. \ Experimental
measurements find a quartet scattering length $^{4}a_{nd}=6.35(2)$~fm
\cite{Dilg:1971}. \ This corresponds with $\kappa\,^{4}a_{nd}=1.47(1)$. \ The
agreement is better when expressed as fraction of the spin-triplet
proton-neutron scattering length, $^{4}a_{nd}/^{3}a_{np}=1.17(1)$. \ 
The 30\% difference between the two values gives an indication 
of higher order effective range effects. A more
detailed calculation including interaction range effects obtains $^{4}%
a_{nd}=6.33(10)$~fm \cite{BvK:1998,BHvK:1998}, in full agreement with
experimental values.

In contrast with the scattering length, there is only one previously reported
determination of the fermion-dimer effective range parameter. \ The correlated
Gaussian expansion calculation in Ref.~\cite{vonStecher:2008} found $\kappa
r_{fd}\approx0.08(1)$. \ Neutron-deuteron scattering data also favors a small
value for $\kappa r_{fd}$. \ However the sign of $\kappa r_{fd}$ has remained
an open question. \ In our continuum calculations presented here, we use the
STM equation to calculate the scattering length and effective range parameter.
\ We find the values $\kappa a_{fd}=1.17907(1)$ and $\kappa r_{fd}=-0.0383(3)$.

Our main focus though is to benchmark lattice calculations of the
fermion-dimer scattering length and effective range parameter. \ For the
lattice calculation we apply the finite-volume phase-shift analysis of
L\"{u}scher. \ We show that finite-volume topological corrections to the dimer
binding energy must be considered in order to obtain accurate results. \ Once
these topological corrections are included in the finite-volume analysis, we
find that the lattice and continuum calculations are in full agreement. \ We 
use two different lattice Hamiltonian formulations, that agree in the continuum 
limit. \ As a final result we find:
\begin{align}
\kappa a_{fd} = 1.174(9),\quad\kappa r_{fd} = -0.029(13).
\end{align}

\section{Notation and Formalism}

Few-body systems of two-component fermions with short-range interactions and large
scattering lengths in comparison to interparticle distances show universal properties. \
Physics in such systems does not depend on the structure of the interactions at short distances. \ 
The problem of three two-component fermions at low energies can be described
by a local quantum field theory whose only interaction term is a two-body
contact interaction. \ In the following, we will always consider equal
mass fermions with mass $m_{\uparrow}=m_{\downarrow}=m$. 
The extension to unequal masses is straightforward.
 \ The free non-relativistic effective Hamiltonian in momentum space can be written as
\begin{equation}
\label{Free-Non-Relativistic}
H_0 = \sum_{i=\uparrow,\downarrow}\int{d^3\vec{p}~\frac{\vec{p}^{\,2}}{2m}~a_i^{\dagger}(\vec{p})a_i(\vec{p})},
\end{equation}
where $a_{i}$ and $a^{\dagger}_{i}$ are annihilation and creation operators.
 \ In position space these operators can be expressed as
\begin{equation}
\label{operators}
a_i(\vec{r}) = \frac{1}{(2\pi)^3}\int{d^3\vec{p}~e^{i\vec{p}\cdot\vec{r}}a_i(\vec{p})},\quad
a_i^{\dagger}(\vec{r}) = \frac{1}{(2\pi)^3}\int{d^3\vec{p}~e^{-i\vec{p}\cdot\vec{r}}a_i^{\dagger}(\vec{p})}.
\end{equation}
Combining Eqs.~(\ref{Free-Non-Relativistic}) and (\ref{operators}) the free 
Hamiltonian in configuration space is then given by
\begin{align}
\label{FreeHamiltonian}
H_0 & = -\frac{1}{2m}\sum_{i=\uparrow,\downarrow}\int{d^3\vec{r}~a_i^{\dagger}(\vec{r})
\vec{\nabla}^2a_i(\vec{r})}\\\nonumber
    & = \frac{1}{2m}\sum_{i=\uparrow,\downarrow}\int{d^3\vec{r}~\big(\vec{\nabla}a_i^{\dagger}
(\vec{r})\big)\big(\vec{\nabla}a_i(\vec{r})\big)}.
\end{align}
Now we introduce an interaction between the fermions via the potential
\begin{equation}
\label{potential}
V(\vec{r},\vec{r}^{\,\prime}) = \frac{1}{2}\sum_{i,j=\uparrow,\downarrow}\int{d^3\vec{r}}
\int{d^3\vec{r}^{\,\prime} :a_i^{\dagger}(\vec{r})a_i(\vec{r})\mathcal{V}(\vec{r}-\vec{r}
^{\,\prime})a_j^{\dagger}(\vec{r}^{\,\prime})a_j(\vec{r}^{\,\prime}):},
\end{equation}
where $: \,\ldots\, :$ denotes normal ordering. 
At low energies, the potential (\ref{potential}) can be 
replaced by a delta-function interaction
\begin{equation}
\label{EffectiveRangeExp}
\mathcal{V}(\vec{r}-\vec{r}^{\,\prime}) = C_0~\delta^{(3)}(\vec{r}-
\vec{r}^{\,\prime}),
\end{equation}
and the the lowest order 
effective Hamiltonian for two-component fermions is
\begin{equation}
H =-\frac{1}{2m}\sum_{i={\uparrow,\downarrow}}\int d^{3}\vec{r}%
~a^{\dagger}_{i}(\vec{r})\vec{\nabla}^{2}a_{i}(\vec{r}) + C_{0}
\int d^{3}\vec{r}~a^{\dagger}_{\uparrow}(\vec{r})a_{\uparrow}
(\vec{r})a^{\dagger}_{\downarrow}(\vec{r})a_{\downarrow}(\vec{r}).
\label{Hamiltonian}
\end{equation}
$C_{0}$ denotes the two-body 
coupling constant and is directly related to the fermion-fermion
scattering length. It is assumed to be negative so that the interaction 
is attractive. 
\ The  exact value of $C_{0}$ depends on the scheme used to regulate the short distance behavior.
 
\ In the next step we consider the two-body and three-body systems of two-component
fermions in Hamiltonian lattice formalism and use the Lanczos method
\cite{Lanczos:1950} to find the lowest eigenvalues. \ Further details of the
Hamiltonian lattice formulation can be found in
\cite{PhysRevC.70.014007,PhysRevC.73.015202,PhysRevLett.98.182501}.
Let $\vec{n}$ denote spatial lattice points on a three-dimensional $L\times
L\times L$ periodic cube. \ We use lattice units where physical quantities are
multiplied by powers of the spatial lattice spacing $a_{latt}$ to make the combination
dimensionless. \ The two-component fermions are labelled as spin-up and
spin-down, the lattice annihilation operators are written as $a_{\uparrow}
(\vec{n})$ and $a_{\downarrow}(\vec{n})$. \ The free non-relativistic Hamiltonian of
two-component fermions with only short-range interaction corresponding to the
Hamiltonian (\ref{FreeHamiltonian}) on the three dimensional lattice is
\begin{equation}
H_{0}=\frac{3}{m}\sum_{\vec{n},i=\uparrow,\downarrow}a_{i}^{\dagger}(\vec
{n})a_{i}(\vec{n})-\frac{1}{2m}\sum_{\hat{\mu}=\hat{1},\hat{2},
\hat{3}}\sum_{\vec{n},i=\uparrow,\downarrow}\big[a_{i}^{\dagger}(\vec{n})a_{i}(\vec
{n}+\hat{\mu})+a_{i}^{\dagger}(\vec{n})a_{i}(\vec{n}-\hat{\mu})\big],
\label{H0}
\end{equation}
where $\hat{\mu}$ is the spatial lattice unit vector. \ We define the
spin-density operators
\begin{align}
\rho_{\uparrow}(\vec{n})  &  =a_{\uparrow}^{\dagger}(\vec{n})a_{\uparrow}%
(\vec{n})\label{5}\\
\rho_{\downarrow}(\vec{n})  &  =a_{\downarrow}^{\dagger}(\vec{n}%
)a_{\downarrow}(\vec{n}),
\end{align}
and consider two different kinds of Hamiltonians. \ In the first Hamiltonian we
have only a single-site interaction. \ This Hamiltonian is
\begin{equation}
H_{1}=H_{0}+C_{1}\sum_{\vec{n}}\rho_{\uparrow}(\vec{n})\rho_{\downarrow}%
(\vec{n}).
\label{H1}
\end{equation}
We consider a second Hamiltonian using a contact interaction as well as
nearest-neighbour interaction terms in order to eliminate the 
two-body effective range parameter,
\begin{align}
H_{2}  &  =H_{0}+C_{2}\sum_{\vec{n}}\rho_{\uparrow}(\vec{n})\rho_{\downarrow
}(\vec{n})\nonumber\label{H2}\\
&  +C_{2}^{\prime}\sum_{\hat{\mu}=\hat{1},\hat{2},\hat{3}}\sum_{\vec{n}%
}\Big[\rho_{\uparrow}(\vec{n})\rho_{\downarrow}(\vec{n}+\hat{\mu}%
)+\rho_{\uparrow}(\vec{n}+\hat{\mu})\rho_{\downarrow}(\vec{n})\Big].
\end{align}
The finite lattice spacing error in these 
two Hamiltonians is of order $a_{latt}^2$.

\begin{figure}[ptb]%
\centering
\includegraphics*[
width=8cm,
angle=-90
]%
{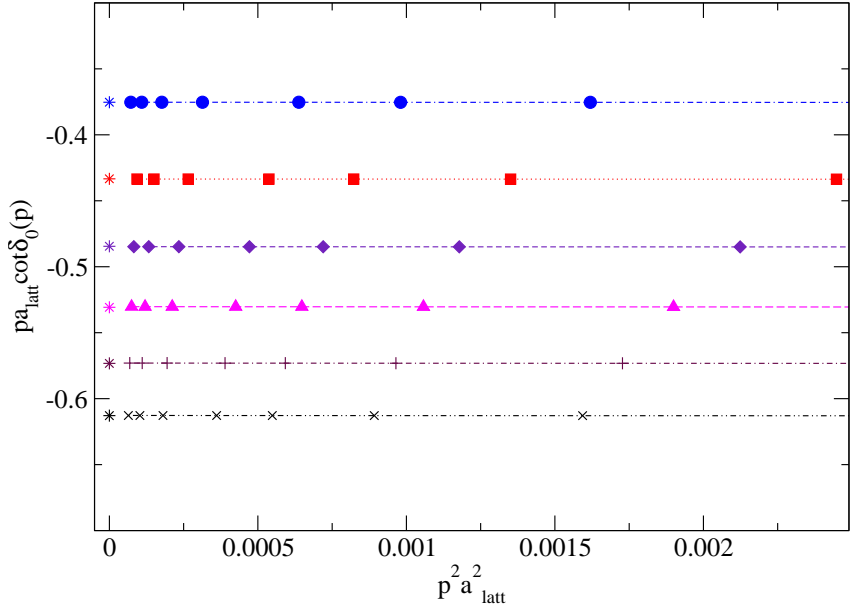}%
\caption{Plot of $p\cot\delta_{0}(p)$ versus $p^{2}$ for the lattice Hamiltonian
$H_{2}$. \scriptsize\ding{108}, \ding{110}, \ding{117}, \ding{115}, \small+, $\times$
\normalsize, represent data points corresponding to the bound states of energies
$-1.5$ MeV, $-2.0$ MeV, $-2.5$~MeV, $-3.0$~MeV, $-3.5$ MeV and $-4.0$ MeV, respectively.
 \small\ding{83} \normalsize denotes $p\cot{\delta_0}(p)$ at the zero momentum limit.}%
\label{pcot_C0C1}
\end{figure}

The next step is to determine the interaction coefficients, $C_{1}$, $C_{2}$ 
and $C_{2}^{\prime}$ using L\"uscher's formula
\cite{Luscher:1985dn,Luscher:1990ux,Beane:2003da}.
\ At present L\"{u}scher's formula is a standard tool in
lattice quantum chromodynamics and in lattice effective field theory.
\ It relates the two-body energy levels 
in a finite volume to the S-wave phase shift.
\begin{equation}
p\cot\delta_{0}(p)=\frac{1}{{\pi}L}S(\eta),\qquad\eta=\Big(\frac{Lp}{2\pi
}\Big)^{2},
\label{pcot}
\end{equation}
where $S(\eta)$ is the three-dimensional zeta-function, $L$ is the length of
the box and $p$ is the center-of-mass momentum. \ The zeta-function in three
dimensions is defined as
\begin{equation}
S(\eta)=\lim_{\Lambda\rightarrow\infty}\Big[\sum_{\vec{n}}\frac{\theta
(\Lambda^{2}-\vec{n}^{2})}{\vec{n}^{2}-\eta}-4\pi\Lambda\Big].
\label{S(eta)}
\end{equation}
For $|\eta|<1$ we can expand $S(\eta)$ in powers of $\eta$,
\begin{align}
S(\eta)  &  =-\frac{1}{\eta}+\lim_{\Lambda\rightarrow\infty}\Big[\sum_{\vec
{n}\neq0}\frac{\theta(\Lambda^{2}-\vec{n}^{2})}{\vec{n}^{2}-\eta}-4\pi
\Lambda\Big],\nonumber\label{expand}\\
&  =-\frac{1}{\eta}+S_{0}+S_{1}\eta+S_{2}\eta^{2}+S_{3}\eta^{3}+\cdots,
\end{align}
where
\begin{equation}
S_{0}=\lim_{\Lambda\rightarrow\infty}\Big[\sum_{\vec{n}\neq0}\frac
{\theta(\Lambda^{2}-\vec{n}^{2})}{\vec{n}^{2}}-4\pi\Lambda\Big],\quad
S_{i}=\sum_{\vec{n}\neq0}\frac{1}{(\vec{n}^{2})^{i+1}}.
\label{coefficients}
\end{equation}
The first few coefficients are
\begin{align}
S_{0}  &  =-8.913631,\quad S_{1}=16.532288,\quad S_{2}=8.401924,\quad
S_{3}=6.945808,\nonumber\label{cofficients}\\
S_{4}  &  =6.426119,\quad S_{5}=6.202149,\quad S_{6}=6.098184,\quad
S_{7}=6.048263.
\end{align}
L\"{u}scher's formula does not include the contribution from higher partial
waves but at asymptotically small momenta we can neglect such corrections.
\ For small momenta we have the effective range expansion,
\begin{equation}
p\cot\delta_{0}(p)\simeq-\frac{1}{a_{s}}+\frac{1}{2}r_{0}~p^{2}+\cdots,
\label{low-energy-exp}
\end{equation}
where $a_{s}$ is the scattering length and $r_{0}$ is the effective range.
Thus the effective range parameters can be extracted from the
finite volume energy levels using Eq.~(\ref{pcot}).
 \ In terms of $\eta$, the energy of the dimer is
\begin{equation}
E=\frac{p^{2}}{m}=\frac{\eta}{m}\Big(\frac{2\pi}{L}\Big)^{2}.
\label{energy}
\end{equation}
The interaction coefficient $C_{1}$ is tuned to construct two-body binding
states (dimers) comprised of one spin-up and one spin-down fermion of energies
$-1.5$ MeV, $-2.0$ MeV, $-2.5$~MeV, $-3.0$ MeV, $-3.5$ MeV and $-4.0$ MeV 
in the large volume ($L=80$) using the Lanczos method. \ In such a large volume the 
finite volume corrections to the dimer binding energy are negligible. \ In our 
calculation we take $m=939$ MeV and $a^{-1}_{latt}=100$ MeV.
\ To find the interaction coefficients of lattice Hamiltonian $H_{2}$ we 
proceed as follows.
\ Setting the effective range to zero requires that the following relation should be 
satisfied near threshold 
\begin{equation}
\frac{1}{\pi L}S(\eta)=p\cot{\delta}_0(p)\simeq-\frac{1}{a_{s}}+O(p^4).
\label{condition}
\end{equation}
The interaction coefficients $C_{2}$ and $C_{2}^{\prime}$ are tuned in order to 
give the binding energies listed above for the ground state in a large 
volume ($L=80$) 
and to fullfil Eq.~(\ref{condition}) for the first excited state.
\ The plot for $p\cot\delta_{0}(p)$ versus
$p^{2}$ for the first excited state is shown in Fig.~\ref{pcot_C0C1}. 
The different values of $p$ are generated by calculating for different
box sizes. \ We note that $p\cot\delta_{0}(p)$ has zero slope near threshold 
since we set the effective range to zero. 
\ Both Hamiltonians reproduce the same continuum limit of fermions with 
attractive zero-range interactions. \ The corresponding 
values for the interaction
coefficients are summarized in Table~\ref{tab1}. 

\begin{table}[t]
\caption{The values of interaction coefficients for the six considered dimers.
All quantities are given in units of the lattice spacing $a_{latt}=(100\mbox{ MeV})^{-1}$.}%
\vspace*{0.2cm}
\centering 
\begin{tabular}{c c c c} 
\hline\hline 
$E^d$\hspace{4ex} & $mC_1$\hspace{4ex} & $mC_2$\hspace{4ex} & $mC^{\prime}_2$\hspace{4ex} \\ [0.5ex]
\hline 
$-0.015$\hspace{4ex} & $-4.51091$\hspace{4ex} & $-4.34554$\hspace{4ex} & $-0.30082$\hspace{4ex} \\%
$-0.020$\hspace{4ex} & $-4.61299$\hspace{4ex} & $-4.45733$\hspace{4ex} & $-0.34273$\hspace{4ex} \\%
$-0.025$\hspace{4ex} & $-4.70675$\hspace{4ex} & $-4.55883$\hspace{4ex} & $-0.34658$\hspace{4ex} \\%
$-0.030$\hspace{4ex} & $-4.79466$\hspace{4ex} & $-4.64749$\hspace{4ex} & $-0.36339$\hspace{4ex} \\%
$-0.035$\hspace{4ex} & $-4.87817$\hspace{4ex} & $-4.73801$\hspace{4ex} & $-0.36452$\hspace{4ex} \\%
$-0.040$\hspace{4ex} & $-4.95823$\hspace{4ex} & $-4.82374$\hspace{4ex} & $-0.36696$\hspace{4ex} \\ [1ex]%
\hline 
\end{tabular}
\label{tab1}
\end{table}

We use the interaction coefficients in Table~\ref{tab1} and diagonalize 
both Hamiltonians (\ref{H1}) and (\ref{H2}) utilizing the Lanczos method 
to determine the ground state energy at rest for the six considered 
dimers in the 
periodic volumes $L^3$ ranging from $L=6$ to $L=17$. We also use the same 
interaction coefficients and diagonalization method to find the ground 
state energy of the fermion-dimer systems. \ These energies are
summarized in the tables in Appendix~A.

Now we turn our attention to the fermion-dimer scattering in a periodic cube. \ It 
is known that there are exponentially small corrections to the scattering energy of
the fermion-dimer system at finite volume due to range effects. \ We can remove this 
error by extrapolation to infinite volume. \ However there is another error which 
is independent of the fermion-dimer scattering process, namely the finite volume 
error in the dimer binding energy. \ 
One might think that this error can be removed by substracting 
the dimer binding energy from the total energy of the 
fermion-dimer scattering system.\ But
this is not quite correct since we calculate the scattering process 
in the center-of-mass frame and therefore the dimer has some recoil momentum.
 \ The corrections to the dimer binding energy in the
moving frame differ from its rest frame due to the topological phases for the
moving dimer in the finite volume \cite{Bour:2011ef,Davoudi:2011md}.
\section{Bound State in a Moving Frame}

L\"uscher derived the finite-volume corrections to the binding energy of 
two-body bound states for interactions with finite range \cite{Luscher:1985dn}.
 \ The shift in the energy of a bound state in a periodic cube at rest is given by
\begin{equation}
\label{mass-shift}
  \Delta E_{\vec{0}}(L) \simeq \sum_{|\vec{n}|=1}\int d^3r
\phi^*_{\infty}(\vec{r})V(\vec{r})\phi_{\infty}(\vec{r}+\vec{n}L),
\end{equation}
where $\phi_{\infty}$ is the infinite-volume wavefunction as a function of 
the relative separation $\vec{r}$ and $V(\vec{r})$ is the interaction potential.
 \ Using a Galilean transformation we can find the wavefunction of the bound state 
in a periodic cube of length $L$ moving with momentum $2\pi\vec{k}/L$ for integer
$\vec{k}$. \ This wavefunction in a periodic cube has a phase dependence 
which can be factorized out,
\begin{equation}
\label{wavefunction}
 \phi_L(\vec{r}+\vec{n}L)=e^{-2i\pi\alpha\vec{k}\cdot\vec{n}}\phi_L(\vec{r}),
\end{equation}
where $\alpha=m_{\uparrow}/(m_{\uparrow}+m_{\downarrow})$ 
for the general case of unequal masses and $\vec{n}$ is an integer. \ 
Each phase twist at the boundaries induces a shift in the binding energy of the dimer.
 \ Using Eqs.~(\ref{mass-shift}) and (\ref{wavefunction}) for the $S$-wave bound state, 
the finite volume correction in a moving frame is
\begin{equation}
\Delta E_{\vec{k}}(L)\approx\tau(\vec{k},\alpha)~\Delta E_{\vec{0}}(L),
\label{Energy-in-boosted-frame}
\end{equation}
where
\begin{equation}
\tau(\vec{k},\alpha)=\frac{1}{3}\sum_{i=1}^{3}\cos(2\pi\alpha k_{i}),
\label{Tau}
\end{equation}
and $\Delta E_{\vec{k}}(L)$ and $\Delta E_{\vec{0}}(L)$ represent finite volume 
correction to the binding energy of the dimer in the moving and rest frame,
respectively.%

These corrections have a universal dependence on momentum determined by the number 
and mass of the constituents. \ In asymptotically large volumes the corrections are 
exponentially small and can be neglected. \ But if the volume is relatively small, 
this shift can be comparable to that of the scattering process of the fermion and dimer. 

In order to calculate finite-volume corrections due to the
binding energy of the dimer in the total scattering energy of the fermion-dimer
system, let $E^{fd}(p,L)$ be the total scattering energy with radial momentum
$p$ and $E_{\vec{k}}^{d}(L)$ the finite volume energy due to binding for the
dimer with momentum $2\pi\vec{k}/L$ in a periodic cube of length $L$. \ 
In the asymptotic limit 
$L\rightarrow\infty$, with $p$ scaling as $1/L$, we can neglect the 
mixing with 
higher-order singular solutions to the Helmholtz equation. \ For S-wave scattering
of a dimer and a fermion with radial momentum $p$ and separation $\vec{r}$ between the 
center of mass of the dimer and the fermion, the position-space scattering wavefunction is
\begin{align}
 \big<\vec{r}~\big|\Psi_p\big>=c\sum_{\vec{k}}\frac{e^{\frac{2i\pi\vec{k}\cdot\vec{r}}{L}}}
  {(2\pi\vec{k}/L)^2-p^2}
\end{align}
with some normalization constant $c$.
\ The total energy $E^{fd}(p,L)$ is given by
\begin{equation}
E^{fd}(p,L)=\frac{\big<\Psi_p\big|H\big|\Psi_p\big>}{\big<\Psi_p\big|\Psi_p\big>}
=\frac{1}{\mathcal{N}}\sum_{\vec{k}}\frac{\frac{p^{2}}{m
}+E_{\vec{k}}^{d}(L)}{(\vec{k}^{2}-\eta)^{2}},
\label{wavefunction-threebody}
\end{equation}
where $\mathcal{N}=\sum_{\vec{k}}(\vec{k}^{2}-\eta)^{-2}$. 
\ The finite-volume correction due to the binding energy of
the dimer in the scattering process is
\begin{equation}
E^{fd}(p,L)-E^{fd}(p,\infty)=\tau_{d}(\eta)~\Delta E_{\vec{0}}^{d}(L),
\label{total-energy}
\end{equation}
where the topological volume factor for $m_{\uparrow}=m_{\downarrow}=m$ is given by
\begin{equation}
\tau_{d}(\eta)=
\frac{1}{\mathcal{N}}\sum_{\vec{k}}\frac
{\tau(\vec{k},\frac{1}{2})}{(\vec{k}^{2}-\eta)^{2}}\,.
\label{topologicalfactor}
\end{equation}
We find $\tau_d(\eta)$ iteratively. \ We determine $p^2$ and $\eta$ by subtracting 
the ground state energy of the dimer in the rest frame from 
the total energy. \ Using this 
$\eta$ and Eq.~(\ref{topologicalfactor}) 
we find $\tau_d(\eta)$ and use it to modify 
the binding energy of the dimer. \ We repeat this process until 
$\eta$ does not change anymore. \ For very large box length ($L=2000$) 
we achieve a fixed point after three iterations.
 \ The ratio of the finite volume corrections in the moving and 
rest frame are collected in Table \ref{tab2}. \ We note that 
this ratio is significantly smaller than one at small volumes. \
For large volumes this ratio can be neglected. \ The binding 
energies of the dimers in the boosted frame are summarized in Appendix~A. \ 
We emphasize that the 
finite-volume correction in Eq.~(\ref{total-energy}) has nothing to do with the 
interaction between dimer and fermion and should therefore be subtracted from the 
total energy before using L\"uscher's scattering relation. \ This subtraction 
reduces systematic errors in lattice calculations.
\ We note that in the total scattering energy there are also corrections
corresponding to the scattering process which we will remove by extrapolation to
the infinite volume and to the continuum limit.%
\begin{table}[t]
\caption{Toplogical factor for six considered dimers. The upper part of table calculated 
by using the ground state energies corresponding to $H_1$ and the lower part by using 
the ground state energies corresponding to $H_2$.}%
\vspace*{0.2cm}
\centering
\scriptsize\begin{tabular}{c c c c c c c c c c c c c}
\hline\hline
$L$\hspace{2ex} & 17\hspace{2ex} & 16\hspace{3ex} & 15\hspace{2ex} & 14\hspace{2ex} &
13\hspace{2ex} & 12\hspace{2ex} & 11\hspace{2ex} & 10\hspace{2ex} & 9\hspace{2ex} &
8\hspace{2ex} & 7\hspace{2ex} & 6\hspace{2ex} \\
$\tau^{1.5}_d$ & 0.84351 & 0.81836 & 0.78862 & 0.75386 & 0.71392 & 0.66915 & 0.62052 & 
0.56968 & - & - & - & - \\%
$\tau^{2.0}_d$ & 0.89079 & 0.87205 & 0.84912 & 0.82108 & 0.78709 & 0.74648 & 0.69916 & 
0.64592 & 0.58872 & - & - & -  \\%
$\tau^{2.5}_d$ & 0.91790 & 0.90351 & 0.88565 & 0.86323 & 0.83517 & 0.80025 & 0.75743 & 
0.70632 & 0.64778 & 0.58452 & - & - \\%
$\tau^{3.0}_d$ & 0.93511 & 0.92349 & 0.90912 & 0.89090 & 0.86769 & 0.83803 & 0.80038 & 
0.75340 & 0.69669 & 0.63176 & 0.56267 & - \\%
$\tau^{3.5}_d$ & 0.94645 & 0.93706 & 0.92520 & 0.91009 & 0.89060 & 0.86532 & 0.83248 & 
0.79022 & 0.73731 & 0.67442 & 0.60554 & - \\%
$\tau^{4.0}_d$ & 0.95466 & 0.94675 & 0.93676 & 0.92396 & 0.90735 & 0.88560 & 0.85681 & 
0.81889 & 0.76973 & 0.70858 & 0.63804 & 0.56535 \\%
\hline
$\tau^{1.5}_d$ & 0.82960 & 0.80247 & 0.77064 & 0.73384 & 0.69221 & 0.64649 & 0.59810 & 
0.54894 & - & - & - & -\\%
$\tau^{2.0}_d$ & 0.88030 & 0.85976 & 0.83471 & 0.80429 & 0.76775 & 0.72470 & 0.67555 & 
0.62176 & 0.56594 & - & - & - \\%
$\tau^{2.5}_d$ & 0.90967 & 0.89377 & 0.87401 & 0.84938 & 0.81873 & 0.78093 & 0.73524 & 
0.68189 & 0.62272 & 0.56130 & - & -\\%
$\tau^{3.0}_d$ & 0.92804 & 0.91529 & 0.89930 & 0.87907 & 0.85338 & 0.82077 & 0.77979 & 
0.72952 & 0.67045 & 0.60542 & 0.53965 & - \\%
$\tau^{3.5}_d$ & 0.94077 & 0.93030 & 0.91709 & 0.90025 & 0.87859 & 0.85060 & 0.81446 & 
0.76847 & 0.71181 & 0.64593 & 0.57567 & - \\%
$\tau^{4.0}_d$ & 0.94988 & 0.94107 & 0.92992 & 0.91564 & 0.89714 & 0.87294 & 0.84112 & 
0.79953 & 0.74636 & 0.68156 & 0.60882 & 0.53618 \\%
\hline
\end{tabular}
\label{tab2}
\end{table}
\begin{figure}[ptb]%
\centering
\includegraphics*[
width=10cm,
angle=270
]%
{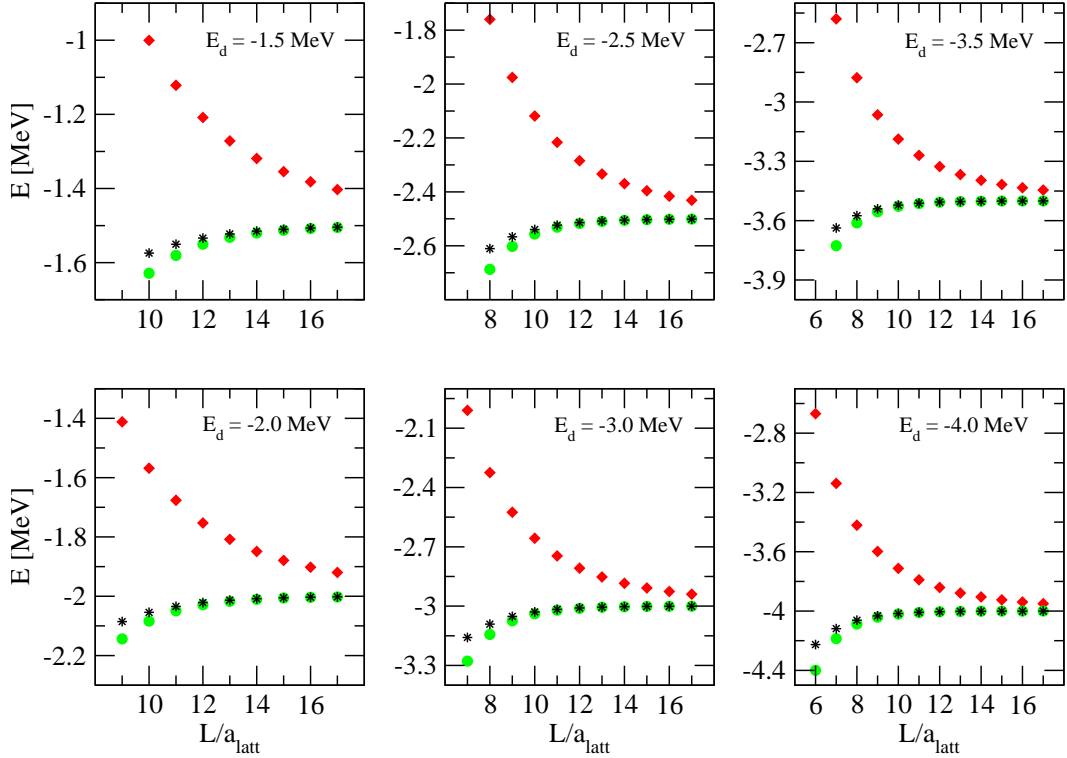}%
\caption{Plots of ground state energies corresponding to the lattice Hamiltonian
$H_1$ versus $L$. Circle and star represent the ground state energies of the dimers
in the rest- and boosted-frame respectively and diamond stands for ground state
energies of the fermion-dimer system.}%
\label{energies_C0}
\end{figure}%

\section{FERMION-DIMER SCATTERING}
\subsection{Lattice Calculation}

\begin{figure}[ptb]%
\centering
\includegraphics*[
width=10cm,
angle=270
]%
{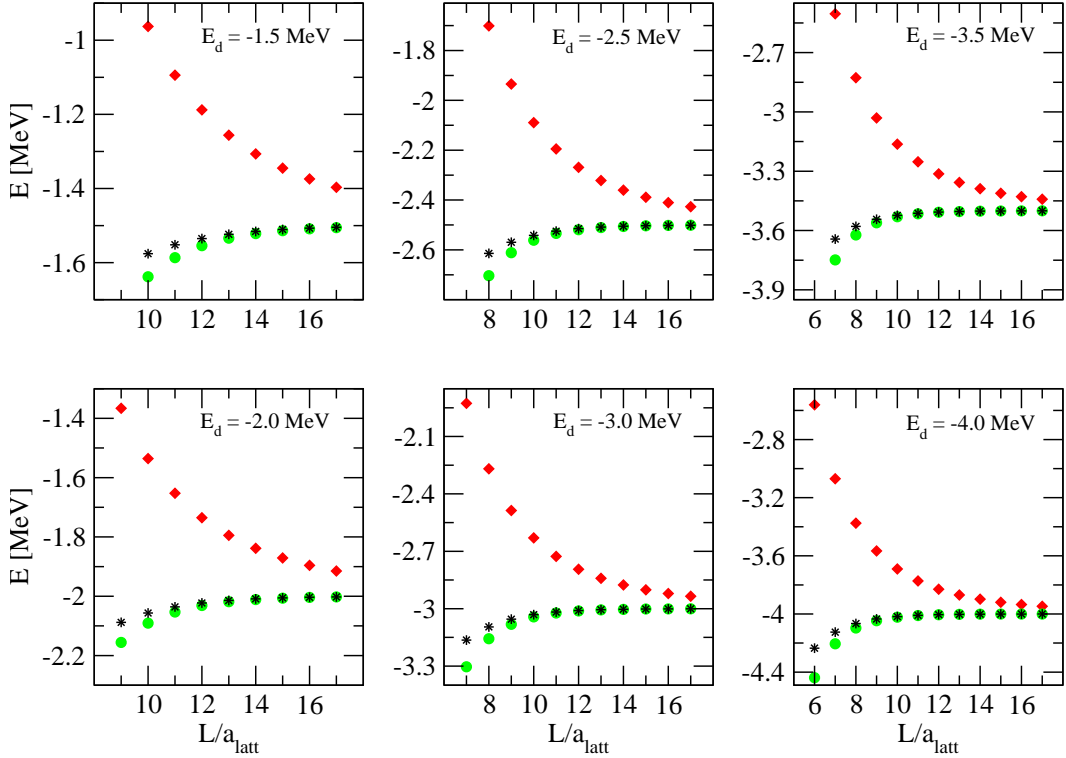}%
\caption{Plots of ground state energies corresponding to the lattice Hamiltonian
$H_2$ versus $L$. Circle and star represent the ground state energies of the dimers
in the rest- and boosted-frame respectively and diamond stands for ground state
energies of the fermion-dimer system.}%
\label{energies_C0C1}
\end{figure}

In order to find the radial momentum, $p$, in the fermion-dimer systems, 
we subtract the binding energies of the dimers in the moving frame from 
the total scattering energies of the fermion-dimer systems.
 \ The ground state energies of the six considered dimers in the rest and 
moving frame and the ground state energies of the
fermion-dimer system for both lattice Hamiltonians, $H_1$ 
and $H_2$, are shown in Figs.~\ref{energies_C0} and \ref{energies_C0C1}.
 \ As discussed above, the difference between the ground state energies of the dimers 
in the rest and moving frame is bigger in small volumes. In this case it is comparable to 
the corrections due to the scattering process.%

\begin{figure}[ptb]%
\centering
\includegraphics*[
height=6.2in,
angle=270
]%
{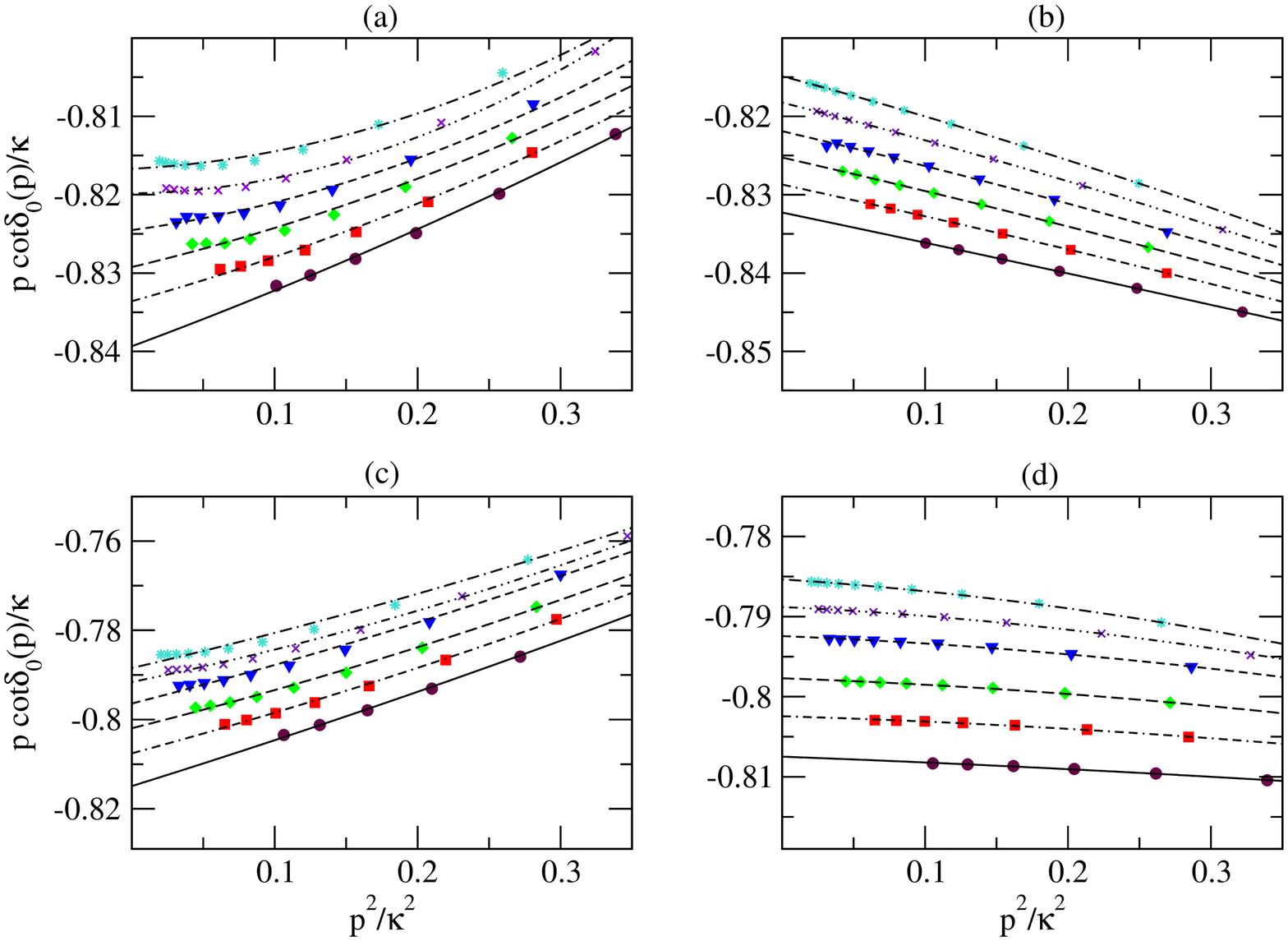}%
\caption{Plots of $p\cot\delta_{0}(p)$ versus $p^{2}$. (a) Naive calculation
for $H_{1}$. (b) Full calculation for $H_{1}$. (c) Naive calculation for
$H_{2}$. (d) Full calculation for $H_{2}$. \large$\bullet$, \scriptsize$\blacksquare$,
\footnotesize$\blacklozenge$, $\blacktriangledown$, \normalsize$\times$ and \small
\ding{83} \normalsize represent data points corresponding to the bound
states of energies $-1.5$ MeV, $-2.0$ MeV, $-2.5$ MeV, $-3.0$ MeV, $-3.5$ MeV
and $-4.0$ MeV respectively.}%
\label{pcot_all}
\end{figure}
The difference $\Delta E(L)$ is the kinetic energy of the fermion-dimer system.
 \ We use this to find the radial momentum $p$. \ In a naive calculation, we 
take $\tau_d(\eta)$ equal to one and thus implicitly 
assume the corrections to the binding energy of the dimer the in rest and moving frame are 
equal. We subtract the binding energy of the dimer in the rest frame from the total energy of 
the fermion-dimer system in order to find $\Delta E(L)$ and eliminate the correction due to 
the dimer binding energy in finite volume,
\begin{equation}
\Delta E^{naive}(L)=\Delta E^{fd}(L)-\Delta E^d_{\vec{0}}(L).
\label{faulty}
\end{equation}
In the full calculation, we subtract the binding energy of the dimer in the boosted 
frame in order to determine $\Delta E(L)$ and eliminate the finite volume correction 
corresponding to the binding energy of the dimer,
\begin{align}
\Delta E^{full}(L) & =\Delta E^{fd}(L)-\Delta E^d_{\vec{k}}(L)\nonumber\\
& =\Delta E^{fd}(L)-\tau_d(\eta)\Delta E^d_{\vec{0}}(L).
\label{full}
\end{align}
After subtracting the corrections corresponding to the binding energy of the
dimer from the total energy we use the L\"{u}scher's formula with this energy and
calculate $p\cot\delta_{0}(p)$ for six different lattice spacings. \ We also
calculate $p\cot\delta_{0}(p)$ for the case in which we subtracted only the
binding energy of the dimer at rest frame from the total energy. \ This simply means
we replace $\tau_{d}(\eta)$ by $1$. \ In  Fig.~\ref{pcot_all}, 
the results are plotted versus $p^{2}$. \ To extrapolate to the infinite volume we fit 
a polynomial of second order to the data points. \ We write this 
results as dimensionless combinations multiplied by powers of the dimer 
binding momentum $\kappa$. \ By comparing the naive calculation plots in  Fig.~\ref{pcot_all}
(a) and (c) with the full calculation plots (b) and (d), we clearly see the effect of the topological 
phase factor. \ This correction is quite large for scattering in smaller volumes.
\ The change in slope in plot (d) compared to plot (b) is expected since we
tuned the effective range of interaction to zero for $H_{2}$. \ From these results, we
determine the low-energy parameters for fermion-dimer scattering and extrapolate
to the continuum limit. 

\subsection{STM Equation}

The STM equation for the S-wave fermion-dimer scattering 
amplitude ${\cal T}_{fd} (k, p; E)$ can be written as \cite{BvK:1998,BHvK:1998}
\begin{eqnarray}
{\cal T}_{fd} (k, p; E) & = & -{8 \kappa \over 3}\, M(k,p;E)
- {2 \over \pi} \int_0^\infty
{dq \, q^2 \, M(q,p;E)\, {\cal T}_{fd} (k, q; E)\over  
-{\kappa} + \sqrt{3q^2/4 -mE -i \epsilon}}
\,,
\label{eq:BHvK}
\end{eqnarray}
where $k$ and $p$ are the incoming and outgoing momenta of the
fermion and dimer in the center-of-mass frame,
$\kappa$ is the binding momentum of the dimer, and
\begin{equation}
E= \frac{3}{4}\frac {p^2}{m}-\frac{\kappa^2}{m}
\end{equation}
is the total energy.
The inhomogeneous term 
\begin{eqnarray}
M(k,p;E)&=& {1 \over 2pk} \ln \left({p^2 + pk + k^2 -mE \over
p^2 - pk + k^2 - mE}\right) \,,
\label{eq:MkpE}
\end{eqnarray}
is given by the S-wave projected one-fermion exchange.
The fermion-dimer scattering phase shifts are
obtained by evaluating Eq.~(\ref{eq:BHvK}) at the on-shell 
point:
\begin{equation}
{\cal T}_{fd} (p, p; E)=\frac{1}{p\cot\delta_0(p)-ip}\,.
\end{equation}
By discretizing the momenta $p$ and $k$
the STM equation (\ref{eq:BHvK}) can be transformed into a matrix equation 
which can be solved numerically. The numerical errors in the
solution of this equation are negligible.
The effective range parameters are extracted from the scattering amplitude
by fitting a polynomial in $p^2$ to $1/{\cal T}_{fd} (p, p; E)+ip$.
Errors are estimated by varying the degree of the fitted 
polynomial. 

\section{Compararison of Results and Discussion}

\begin{figure}[ptb]%
\centering
\includegraphics*[
height=6.2in,
angle=270
]%
{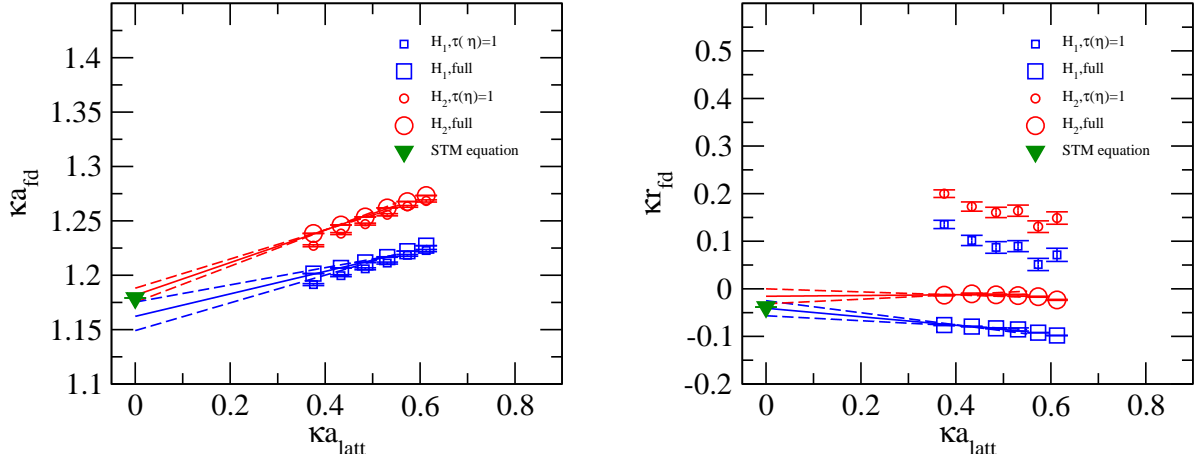}
\caption{Left: Lattice results and continuum extrapolation with error
estimates for the fermion-dimer scattering length. Right: Lattice results and
continuum extrapolation with error estimates for the fermion-dimer effective range.
For comparison we show the continuum result obtain via the
Skorniakov-Ter-Martirosian equation. The error bands show statistical errors 
corresponding to the fitting procedure shown in Fig.~\ref{pcot_all}}%
\label{parameter}
\end{figure}

From the lattice calculation of the phase shifts in  Fig.~\ref{pcot_all} we can extract
the effective range parameters.\ Our results for the scattering length, $a_{fd}$, and
the effective range parameter, $r_{fd}$, are shown in Fig.~\ref{parameter}.
%
We analyze only the plots in Fig.~\ref{pcot_all} (b) and (d) which contain the full calculations 
corresponding to $H_1$ and $H_2$, respectively. \ By fitting a polynomial 
of second order to each set of data we find a scattering length and a 
effective range in infinite volume for both lattice Hamiltonians. 
\ These data points are plotted in Fig~\ref{parameter}. \ In order to 
extrapolate to the continuum limit $a_{latt}\to 0$, we use a linear function. 
\ The results for the low-energy parameters that we get for these two
independent representations of the lattice Hamiltonians are
\begin{align}
\kappa a_{fd}  &  =1.162(13),\quad\kappa r_{fd}=-0.041(16)\quad\text{for}%
~H_{1}\label{results1}\\
\kappa a_{fd}  &  =1.181(7),\quad\kappa r_{fd}=-0.016(16)\quad\text{for}%
~H_{2}.\label{results2}
\end{align}

 \ To extrapolate to the continuum limit in the lattice Hamiltonian calculations we 
used only the data points corresponding to the four smallest lattice spacings.
 \ For the other data points, the Compton wavelength of the bound state 
is comparable to the lattice spacing. \ We estimate the systematic errors in 
the continuum extrapolation of the fermion-dimer scattering length and effective 
range by extrapolation to the continuum limit using only the first two data points 
and taking the interval between these extrapolation values and the central values 
obtained using all four data points as the systematic errors. \ The agreement 
between these two independent calculations is consistent with our estimate of 
the systematic errors. \ As we see from Fig.~\ref{parameter} the inclusion 
of the topological volume factor $\tau_d(\eta)$ improves the accuracy, especially 
in the calculation of the effective range parameter. \ With a very conservative 
estimation of the systematic error we are able to say that the value of the 
the fermion-dimer scattering length in units of the dimer binding momentum
is in between $1.149$ and $1.188$. \ The value of fermion-dimer effective range 
in units of the dimer binding momentum is between zero and $-0.057$. 
 \ Our final result is given by the weighted averages of 
the values in Eq.~(\ref{results1}) and (\ref{results2}):
\begin{align}
\kappa a_{fd} = 1.174(9),\quad\kappa r_{fd} = -0.029(13).
\label{resultF}
\end{align}
In calculating the average,
we assumed that the statistical probability distribution of the measured
variables are Gaussian and independent of each other. \ Using standard error 
propagation, we find the uncertainty in the average values.
\ Our results (\ref{resultF})
are in excellent agreement with the continuum calculation using 
the STM integral equation (\ref{eq:BHvK}):
\begin{equation}
\kappa a_{fd}=1.17907(1),\quad\kappa r_{fd}=-0.0383(3).
\label{STM}
\end{equation}

\section{Outlook}

We have presented benchmark calculations for fermion-dimer scattering using a
Hamiltonian lattice formalism and in the continuum using the STM integral
equation. \ We obtain excellent agreement between both approaches. \ The
finite-volume lattice methods presented here can be applied to \textit{ab
initio} calculations for elastic scattering of nuclei, cold atoms, and
hadronic molecules. \ Of particular interest in nuclear physics are
calculations of the low-energy scattering of neutrons upon nuclei. \ Soft
neutron scattering upon nuclei is relevant to the design of container
materials used in ultracold neutron experiments. \ They are also an important
probe of the properties of nuclei surrounded by a dilute superfluid neutron
gas. \ It is widely believed that this physical situation with nuclei in a
neutron gas is realized in the inner crust of neutron stars.

Calculations of deuteron-neutron scattering in the spin-quartet channel should
yield similar results to the idealized zero-range limit presented here.
\ There will, however, be small corrections due to the range of the interactions
as well as spin-dependent forces. \ Our methods can also be applied to
neutron-deuteron scattering in the spin-doublet channel. \ Here there is an
interesting connection with Efimov trimer physics in a finite volume
\cite{Kreuzer:2008bi,Kreuzer:2010ti,Kreuzer:2012sr}. \ As a first step in this
direction, boson-dimer scattering for three identical bosons in the zero-range
limit is currently being investigated \cite{Rokash:2012}.

There are also some very useful applications of the topological volume factor
for binding energy calculations. \ By choosing different values of the center
of mass motion, one can make the sum over topological volume factors vanish.
\ This can be used to remove the leading and even subleading finite-volume
corrections to the binding energy of two-body bound states. \ This technique
is being pursued in a number of recent lattice QCD studies for the deuteron
and other dibaryonic systems\ \cite{Davoudi:2011md,Beane:2011iw}.

\acknowledgments
We thank D\"orte Blume and Martin Savage for useful discussions. \ Partial financial 
support from the Deutsche Forschungsgemeinschaft (SFB/TR 16), Helmholtz Association
(contract number VH-VI-231), BMBF (grant 06BN9006), and U.S. Department of Energy (DE-FG02-
03ER41260) are acknowledged. \ This work was further supported by the EU HadronPhysics3 
project "Study of strongly interacting matter".

\appendix 
\section{Ground State Energies}
\begin{table}[htb!]
\caption{Ground state energy of the dimers in the rest frame calculated by using $H_1$.}%
\centering
\scriptsize\begin{tabular}{c c c c c c c c c c c c c}
\hline\hline
$L$\hspace{2ex} & 17\hspace{2ex} & 16\hspace{3ex} & 15\hspace{2ex} & 14\hspace{2ex} &
13\hspace{2ex} & 12\hspace{2ex} & 11\hspace{2ex} & 10\hspace{2ex} & 9\hspace{2ex} & 
8\hspace{2ex} & 7\hspace{2ex} & 6\hspace{2ex} \\
$E^{1.5}_{\vec0}$ & -1.5051 & -1.5080 & -1.5126 & -1.5200 & -1.5316 & -1.5504 & -1.5804 &
 -1.6287 & - & - & - & - \\%
$E^{2.0}_{\vec0}$ & -2.0022 & -2.0036 & -2.0060 & -2.0101 & -2.0170 & -2.0287 & -2.0489 & 
-2.0838 & -2.1439 & - & - & -\\%
$E^{2.5}_{\vec0}$ & -2.5010 & -2.5018 & -2.5031 & -2.50550 & -2.5097 & -2.5173 & -2.5311& 
-2.5562 & -2.6024& -2.6874 & - & - \\%
$E^{3.0}_{\vec0}$ & -3.0005 & -3.0009 & -3.0017 & -3.0032 & -3.0059 & -3.0109 & -3.0205 & 
-3.0388 & -3.0743 & -3.1434 & -3.2782 & - \\%
$E^{3.5}_{\vec0}$ & -3.5003 & -3.5005 & -3.5010 & -3.5019 & -3.5037 & -3.5071 & -3.5139 & 
-3.5275 & -3.5550 & -3.6112 & -3.7266 & - \\%
$E^{4.0}_{\vec0}$ & -4.0002 & -4.0003 & -4.0006 & -4.0012 & -4.0024 & -4.0048 & -4.0097 & 
-4.0200 & -4.0415 & -4.0874 & -4.1861 & -4.3994 \\%
\hline
\end{tabular}
\label{tab3}
\end{table}
\begin{table}[htb!]
\caption{Ground state energy of the dimers in the rest frame calculated by using $H_2$.}%
\centering
\scriptsize\begin{tabular}{c c c c c c c c c c c c c}
\hline\hline
$L$\hspace{2ex} & 17\hspace{2ex} & 16\hspace{3ex} & 15\hspace{2ex} & 14\hspace{2ex} &
13\hspace{2ex} & 12\hspace{2ex} & 11\hspace{2ex} & 10\hspace{2ex} & 9\hspace{2ex} & 
8\hspace{2ex} & 7\hspace{2ex} & 6\hspace{2ex} \\
$E^{1.5}_{\vec0}$ & -1.5055 & -1.5087 & -1.5136 & -1.5215 & -1.5341 & -1.5543 & -1.5865 & 
-1.6382 & - & - & - & - \\%
$E^{2.0}_{\vec0}$ & -2.0024 & -2.0040 & -2.0066 & -2.0110 & -2.0185 & -2.0313 & -2.0532 & 
-2.0908 & -2.1556 & - & - & - \\%
$E^{2.5}_{\vec0}$ & -2.5011 & -2.5020 & -2.5034 & -2.5060 & -2.5106 & -2.5189 & -2.5339 & 
-2.5613 & -2.6115 & -2.7035 & - & - \\%
$E^{3.0}_{\vec0}$ & -3.0006 & -3.0011 & -3.0020 & -3.0035 & -3.0065 & -3.0121 & -3.0226 & 
-3.0428 & -3.0817 & -3.1573 & -3.3039 & - \\%
$E^{3.5}_{\vec0}$ & -3.5003 & -3.5006 & -3.5011 & -3.5021 & -3.5041 & -3.5079 & -3.5154 & 
-3.5304 & -3.5608 & -3.6225 & -3.7487 & - \\%
$E^{4.0}_{\vec0}$ & -4.0004 & -4.0005 & -4.0009 & -4.0015 & -4.0028 & -4.0055 & -4.0110 & 
-4.0223 & -4.0462 & -4.0968 & -4.2052 & -4.4384 \\%
\hline
\end{tabular}
\label{tab4}
\end{table}
\begin{table}[htb!]
\caption{Ground state energy of fermion-dimer systems calculated by using $H_1$.}%
\centering
\scriptsize\begin{tabular}{c c c c c c c c c c c c c}
\hline\hline
$L$\hspace{2ex} & 17\hspace{2ex} & 16\hspace{3ex} & 15\hspace{2ex} & 14\hspace{2ex} &
13\hspace{2ex} & 12\hspace{2ex} & 11\hspace{2ex} & 10\hspace{2ex} & 9\hspace{2ex} &
8\hspace{2ex} & 7\hspace{2ex} & 6\hspace{2ex} \\%
$E^{fd}_{1.5}$ & -1.4029 & -1.3817 & -1.3545 & -1.3189 & -1.2718 & -1.2084 & -1.1216 & -1.0003 & 
- & - & - & -\\%
$E^{fd}_{2.0}$ & -1.9199 & -1.9022 & -1.8791 & -1.8487 & -1.8081 & -1.7529 & -1.6765 & -1.5684 & 
-1.4118 & - & - & - \\%
$E^{fd}_{2.5}$ & -2.4313 & -2.4159 & -2.3959 & -2.3693 & -2.3335 & -2.2845 & -2.2161 & -2.1184 & 
-1.9754 & -1.7601 & - & - \\%
$E^{fd}_{3.0}$ & -2.9395 & -2.9258 & -2.9080 & -2.8843 & -2.8523 & -2.8081 & -2.7461 & -2.6568 & 
-2.5250 & -2.3246 & -2.0095 & - \\%
$E^{fd}_{3.5}$ & -3.4455 & -3.4332 & -3.4172 & -3.3958 & -3.3667 & -3.3265 & -3.2696 & -3.1873 & 
-3.0649 & -2.8773 & -2.5796 & - \\%
$E^{fd}_{4.0}$ & -3.9502 & -3.9390 & -3.9244 & -3.9048 & -3.8782 & -3.8412 & -3.7886 & -3.7122 & 
-3.5978 & -3.4212 & -3.1388 & -2.6686 \\%
\hline
\end{tabular}
\label{tab5}
\end{table}
\begin{table}[htb!]
\caption{Ground state energies of fermion-dimer systems calculated by using $H_2$.}%
\centering
\scriptsize\begin{tabular}{c c c c c c c c c c c c c}
\hline\hline
$L$\hspace{2ex} & 17\hspace{2ex} & 16\hspace{3ex} & 15\hspace{2ex} & 14\hspace{2ex} &
13\hspace{2ex} & 12\hspace{2ex} & 11\hspace{2ex} & 10\hspace{2ex} & 9\hspace{2ex} &
8\hspace{2ex} & 7\hspace{2ex} & 6\hspace{2ex} \\%
$E^{fd}_{1.5}$ & -1.3967 & -1.3741 & -1.3449 & -1.3067 & -1.2561 & -1.1878 & -1.0941 & -0.9629 & 
- & - & - & - \\%
$E^{fd}_{2.0}$ & -1.9148 & -1.8958 & -1.8710 & -1.8384 & -1.7947 & -1.7353 & -1.6528 & -1.5359 & 
-1.3663 & - & - & - \\%
$E^{fd}_{2.5}$ & -2.4267 & -2.4103 & -2.3887 & -2.3602 & -2.3217 & -2.2689 & -2.1951 & -2.0895 & 
-1.9346 & -1.7011 & - & - \\%
$E^{fd}_{3.0}$ & -2.9353 & -2.9207 & -2.9015 & -2.8760 & -2.8414 & -2.7938 & -2.7266 & -2.6299 & 
-2.4868 & -2.2691 & -1.9264 & - \\%
$E^{fd}_{3.5}$ & -3.4418 & -3.4286 & -3.4114 & -3.3883 & -3.3570 & -3.3136 & -3.2522 & -3.1631 & 
-3.0304 & -2.8269 & -2.5039 & - \\%
$E^{fd}_{4.0}$ & -3.9470 & -3.9351 & -3.9193 & -3.8983 & -3.8696 & -3.8297 & -3.7730 & -3.6903 & 
-3.5665 & -3.3753 & -3.0694 & -2.5611 \\%
\hline
\end{tabular}
\label{tab6}
\end{table}
\begin{table}[htb!]
\caption{Ground state energy of dimers in the boosted frame calculated by using $H_1$.}%
\centering
\scriptsize\begin{tabular}{c c c c c c c c c c c c c}
\hline\hline
$L$\hspace{2ex} & 17\hspace{2ex} & 16\hspace{3ex} & 15\hspace{2ex} & 14\hspace{2ex} &
13\hspace{2ex} & 12\hspace{2ex} & 11\hspace{2ex} & 10\hspace{2ex} & 9\hspace{2ex} &
8\hspace{2ex} & 7\hspace{2ex} & 6\hspace{2ex} \\%
$E^{1.5}_{\vec{k}}$ & -1.5043 & -1.5066 & -1.5100 & -1.5150 & -1.5226 & -1.5338 & -1.5502 & 
-1.5741 & - & - & - & - \\%
$E^{2.0}_{\vec{k}}$ & -2.0020 & -2.0032 & -2.0051 & -2.0083 & -2.0134 & -2.0215 & -2.0343 & 
-2.0543 & -2.0855 & - & - & - \\%
$E^{2.5}_{\vec{k}}$ & -2.5009 & -2.5016 & -2.5028 & -2.5047 & -2.5081 & -2.5138 & -2.5235 & 
-2.5398 & -2.5666 & -2.6106 & - & - \\%
$E^{3.0}_{\vec{k}}$ & -3.0005 & -3.0009 & -3.0016 & -3.0028 & -3.0051 & -3.0091 & -3.0164 & 
-3.0293 & -3.0519 & -3.0911 & -3.1586 & - \\%
$E^{3.5}_{\vec{k}}$ & -3.5003 & -3.5005 & -3.5009 & -3.5017 & -3.5033 & -3.5062 & -3.5116 & 
-3.5218 & -3.5406 & -3.5750 & -3.6372 & - \\%
$E^{4.0}_{\vec{k}}$ & -4.0001 & -4.0003 & -4.0006 & -4.0011 & -4.0022 & -4.0042 & -4.0083 & 
-4.0164 & -4.0319 & -4.0619 & -4.1187 & -4.2258\\%
\hline
\end{tabular}
\label{tab7}
\end{table}
\begin{table}[htb!]
\caption{Ground state energy of dimers in the boosted frame calculated by using $H_2$.}%
\centering
\scriptsize\begin{tabular}{c c c c c c c c c c c c c}
\hline\hline
$L$\hspace{2ex} & 17\hspace{2ex} & 16\hspace{3ex} & 15\hspace{2ex} & 14\hspace{2ex} &
13\hspace{2ex} & 12\hspace{2ex} & 11\hspace{2ex} & 10\hspace{2ex} & 9\hspace{2ex} &
8\hspace{2ex} & 7\hspace{2ex} & 6\hspace{2ex} \\%
$E^{1.5}_{\vec{k}}$ & -1.5046 & -1.5070 & -1.5105 & -1.5158 & -1.5236 & -1.5351 & -1.5517 & 
-1.5758 & - & - & - & - \\%
$E^{2.0}_{\vec{k}}$ & -2.0022 & -2.0034 & -2.0055 & -2.0089 & -2.0142 & -2.0227 & -2.0359 & 
-2.0565 & -2.0881 & - & - & - \\%
$E^{2.5}_{\vec{k}}$ & -2.5010 & -2.5018 & -2.5030 & -2.5051 & -2.5087 & -2.5148 & -2.5250 & 
-2.5418 & -2.5694 & -2.6142 & - & - \\%
$E^{3.0}_{\vec{k}}$ & -3.0006 & -3.0010 & -3.0018 & -3.0031 & -3.0055 & -3.0099 & -3.0176 & 
-3.0312 & -3.0548 & -3.0952 & -3.1640 & - \\%
$E^{3.5}_{\vec{k}}$ & -3.5003 & -3.5005 & -3.5010 & -3.5019 & -3.5036 & -3.5067 & -3.5125 & 
-3.5234 & -3.5608 & -3.6225 & -3.7487 & - \\%
$E^{4.0}_{\vec{k}}$ & -4.0003 & -4.0005 & -4.0008 & -4.0014 & -4.0025 & -4.0048 & -4.0092 & 
-4.0178 & -4.0344 & -4.0660 & -4.1249 & -4.2350 \\%
\hline
\end{tabular}
\label{tab8}
\end{table}
\newpage
\phantomsection


\end{document}